\documentclass[graybox]{svmult}

\usepackage{mathptmx}       % selects Times Roman as basic font
\usepackage{helvet}         % selects Helvetica as sans-serif font
\usepackage{courier}        % selects Courier as typewriter font
\usepackage{type1cm}        % activate if the above 3 fonts are
\usepackage{makeidx}         % allows index generation
\usepackage{graphicx}        % standard LaTeX graphics tool
\usepackage{multicol}        % used for the two-column index
\usepackage[bottom]{footmisc}% places footnotes at page bottom

\usepackage{amssymb,amsmath}

\newcommand{\beq}{\begin{equation}}
\newcommand{\beqn}{\begin{eqnarray}}
\newcommand{\eeq}{\end{equation}}
\newcommand{\eeqn}{\end{eqnarray}}

\begin{document}

\title*{Nonminimal Couplings in the Early Universe: Multifield Models of Inflation and the Latest Observations}

\titlerunning{Nonminimal Couplings in the Early Universe}
\author{David I. Kaiser}
\institute{Center for Theoretical Physics, Department of Physics, Massachusetts Institute of Technology, 77 Massachusetts Avenue, Cambridge, MA 02139 USA.  \email{dikaiser@mit.edu} }
\maketitle

\abstract{Models of cosmic inflation suggest that our universe underwent an early phase of accelerated expansion, driven by the dynamics of one or more scalar fields. Inflationary models make specific, quantitative predictions for several observable quantities, including particular patterns of temperature anistropies in the cosmic microwave background radiation. Realistic models of high-energy physics include many scalar fields at high energies. Moreover, we may expect these fields to have nonminimal couplings to the spacetime curvature. Such couplings are quite generic, arising as renormalization counterterms when quantizing scalar fields in curved spacetime. In this chapter I review recent research on a general class of multifield inflationary models with nonminimal couplings. Models in this class exhibit a strong attractor behavior: across a wide range of couplings and initial conditions, the fields evolve along a single-field trajectory for most of inflation. Across large regions of phase space and parameter space, therefore, models in this  general class yield robust predictions for observable quantities that fall squarely within the ``sweet spot" of recent observations.\footnote{
Preprint MIT-CTP-4740. Published in {\it At the Frontier of Spacetime: Scalar-Tensor Theory, Bell's Inequality, Mach's Principle, Exotic Smoothness}, ed. T. Asselmeyer-Maluga (Springer, 2016), pp. 41-57, in honor of Carl Brans's 80th birthday.} }

\section{Introduction}
\label{sec:1}

I first met Carl Brans about twenty years ago, in the mid-1990s, when I was a graduate student. Carl invited me to visit him at Loyola University in New Orleans, and he and his wife Anna kindly hosted me in their beautiful home. Our first meeting has always stood out in my mind: Carl picked me up at the airport, drove me straight to his office, and handed me a piece of chalk. I was to give him a lecture, right there at the blackboard, about cosmic inflation. I launched in, as best I could, and after a fun discussion Carl announced that it was time to pause and get some seafood gumbo; after all, we were in New Orleans. Ever since my first visit, I have found it terrifically inspiring to talk with Carl and to try to sharpen my own ideas in the face of his excellent questions, which he has always delivered in a gentle and encouraging way.

Carl pursued what has become known as the ``Brans-Dicke" theory of gravitation for his Ph.D. dissertation at Princeton, working closely with his advisor Robert Dicke \cite{BransDiss,BD61,Brans62}. Previous physicists had explored various ideas for scalar-tensor theories of gravity, including Pascual Jordan's well-known work, though none of the prior efforts had nearly the same galvanizing influence on the physics community as the Brans-Dicke work \cite{Willbook,Norton92,KaiserSciAm,BransEinstein,Goenner2012}. Brans and Dicke were motivated to try to incorporate Mach's principle in a relativistic theory of gravitation more consistently than Einstein had done in his general theory of relativity.\footnote{On Einstein's changing considerations of Mach's principle, see \cite{Janssen} and references therein.}

The key insight in Brans and Dicke's work was to couple a scalar field directly to the Ricci spacetime curvature scalar in the action, thereby replacing Newton's constant, $G$, with an effective strength of gravity that could vary over space and time. Since Brans and Dicke introduced their formative work, several distinct theoretical motivations have emerged for such nonminimal couplings, beyond consideration of Mach's principle, including everything from dimensional compactification of higher-dimensional theories to effective couplings in supergravity and beyond. (For recent discussions, see \cite{FujiiMaeda,Faraoni04,CapoFaraoni2011,Nojiri}.) 

Perhaps the most mundane motivation for such nonminimal couplings today --- but for me, the most compelling --- is that nonminimal couplings arise as necessary counterterms when quantizing a self-interacting scalar field in curved spacetime. Even if the bare coupling is set to zero, quantum corrections will induce a nonzero coupling \cite{CCJ,Bunch,BirrellDavies,Odintsov90,Buchbinder,Parker,Markkanen}. Moreover, the nonminimal coupling typically rises with energy scale under renormalization-group flow, with no ultraviolet fixed point \cite{Buchbinder}. It therefore makes sense to consider models with sizable nonminimal couplings at high energies, at or above the GUT scale --- and hence to consider nonminimal couplings when thinking about the early universe.

\section{Nonminimal Couplings and Inflation}
\label{sec:2}

Models of cosmic inflation suggest that our observable universe underwent an early phase of accelerated expansion, driven by the dynamics of one or more scalar fields \cite{Guth81}. (For reviews, see \cite{BTW,GuthKaiser05}.) There is by now a long history of building models of early-universe inflation incorporating nonmiminal couplings. Early models such as ``induced-gravity inflation" \cite{IgI}, for example, built directly on work by Lee Smolin \cite{Smolin79} and Anthony Zee \cite{Zee79}, who had aimed to combine Brans-Dicke gravitation with a Higgs-like spontaneous symmetry breaking potential, in order to account for why the strength of gravity is so much weaker than the other fundamental forces. ``Extended inflation" \cite{LaSteinhardt} likewise combined a Brans-Dicke field with a simple potential to drive accelerated expansion. Others considered more general nonmiminal couplings, in which the effective gravitational coupling $G_{\rm eff}$ arose as a combination of a bare coupling constant plus contributions from a scalar field coupled to the Ricci curvature scalar \cite{NMInflation}. Among the most prominent recent examples is ``Higgs inflation" \cite{HiggsInflation}. In such models, the scalar field is expected to settle into a minimum of its potential near the end of inflation, leading to an effectively constant gravitational coupling for most of cosmic history. Hence such models present no tension with Solar System constraints on scalar-tensor gravity.

Realistic models of particle physics, relevant for inflationary energy scales, include many scalar fields \cite{LythRiotto}. The renormalization arguments alone suggest that each of these scalar fields should have a nonminimal coupling. So together with several students and collaborators, I have enjoyed exploring in recent years multifield models of inflation that incorporate nonminimal couplings \cite{KT,KMS,GKS,KS,SSK,DeCross}.

The action for the original Brans-Dicke theory may be written
%%%%%%
\beq
S_{BD} = \int d^4 x \sqrt{-\tilde{g} } \left[ \Phi \tilde{R} - \frac{ \omega}{\Phi} \tilde{g}^{\mu\nu} \partial_\mu \Phi \partial_\nu \Phi \right] ,
\label{SBD}
\eeq
where $\omega$ is a dimensionless constant and $\tilde{g}_{\mu\nu} (x)$ is the spacetime metric. (Greek letters label spacetime indices, $\mu, \nu = 0, 1, 2, 3$.) In $(3 + 1)$ spacetime dimensions, the Brans-Dicke field $\Phi$ has dimensions $({\it mass})^2$. Since high-energy theorists typically consider scalar fields that have dimension ${\it mass}$ in $(3 + 1)$ spacetime dimensions, we may rescale the Brans-Dicke field as $\Phi \rightarrow \phi^2 / (8 \omega)$. In terms of the rescaled field $\phi$, the action of Eq. (\ref{SBD}) may be written
%%%%%
\beq
S_{BD} = \int d^4 x \sqrt{- \tilde{g}} \left[ f_{BD} (\phi) \tilde{R} - \frac{1}{2} \tilde{g}^{\mu\nu} \partial_\mu \phi \partial_\nu \phi \right] .
\label{SBD2}
\eeq
The nonminimal coupling function takes the form
%%%%%%
\beq
f_{BD} (\phi) = \frac{1}{2} \xi \phi^2 ,
\label{fBD}
\eeq
where the dimensionless coupling constant $\xi$ is related to the original Brans-Dicke parameter as $\xi = 1/ (8 \omega)$. Such a quadratic term is precisely the form in which quantum corrections arise for scalar fields in curved spacetime, and hence the form that appropriate counterterms must assume \cite{CCJ,Bunch,BirrellDavies,Odintsov90,Buchbinder,Parker,Markkanen}.

In Brans and Dicke's original formulation, the local strength of gravity, $G_{\rm eff} (x)$, varies with the field $\phi (x)$: $G_{\rm eff} (x) = 1 / (8 \pi \xi \phi^2)$. One may generalize such a coupling to include a bare (constant) mass, $M_0$, within the function $f (\phi)$:
%%%%%%
\beq
f (\phi) = \frac{1}{2} \left[ M_0^2 + \xi \phi^2 \right] ,
\label{fsingle}
\eeq
with $(16\pi G_{\rm eff})^{-1} = f (\phi)$. And this form, in turn, may be generalized to models with $N$ scalar fields:
%%%%%
\beq
f (\phi^I) = \frac{1}{2} \left[ M_0^2 + \sum_{I = 1}^N \xi_I \left( \phi^I \right)^2 \right] .
\label{fmulti}
\eeq
We therefore consider models for which the action may be written
%%%%%%
\beq
S = \int d^4 x \sqrt{- \tilde{g} } \left[ f (\phi^I ) \tilde{R} - \frac{1}{2} \delta_{IJ} \tilde{g}^{\mu\nu} \partial_\mu \phi^I \partial_\nu \phi^J - \tilde{V} (\phi^I ) \right] .
\label{SJ}
\eeq
Here capital Latin letters label field-space indices, $I, J = 1 , 2, ... , N$, and tildes denote quantites in the so-called Jordan frame, in which the nonminimal couplings, $ f (\phi^I ) \tilde{R}$, remain explicit in the action.

Because we are interested in comparing predictions from this family of models with recent astrophysical observations --- especially high-precision measurements of the cosmic microwave background radiation (CMB) --- it is convenient to work in the so-called Einstein frame, for which physicists have established a powerful gauge-invariant formalism for treating gravitational perturbations.\footnote{We have bracketed, for now, the important and rather subtle question of whether there remains any significant ``frame dependence" for predictions from such multifield models. It seems clear that one may map predictions for observables from one frame to another in the case of single-field models \cite{Faraoni04,CapoFaraoni2011}. But making that mapping between frames in the presence of entropy (or isocurvature) perturbations --- which can only arise in multifield models --- seems to raise new subtleties \cite{FrameDep}.} (For reviews, see \cite{BTW,GaugeInvariant}.) 

In order to bring the gravitational portion of the action of Eq. (\ref{SJ}) to the familiar Einstein-Hilbert form, we perform a conformal transformation, much as Dicke described early in the study of Brans-Dicke gravitation \cite{Dickeconf}. We rescale the spacetime metric tensor, $\tilde{g}_{\mu\nu} (x) \rightarrow g_{\mu\nu} (x) = \Omega^2 (x) \tilde{g}_{\mu\nu} (x)$. The conformal factor $\Omega^2 (x)$ is positive definite and is related to the nonminimal coupling function that appears in Eq. (\ref{SJ}) as
%%%%%%
\beq
\Omega^2 (x) = \frac{2}{M_{\rm pl}^2} f (\phi^I (x)) ,
\label{Omega}
\eeq
where $M_{\rm pl} \equiv 1 / \sqrt{ 8 \pi G} = 2.43 \times 10^{18}$ GeV is the reduced Planck mass, related to Newton's gravitational constant, $G$.
Upon performing this conformal transformation, the action of Eq. (\ref{SJ}) is transformed to \cite{DKconf}
%%%%%%%%
\beq
S = \int d^4 x \sqrt{-g} \left[ \frac{ M_{\rm pl}^2}{2} R - \frac{1}{2} {\cal G}_{IJ} (\phi^K ) g^{\mu\nu} \partial_\mu \phi^I \partial_\nu \phi^J - V (\phi^I) \right] .
\label{SE}
\eeq
The conformal transformation induces a field-space manifold whose metric, in the Einstein frame, is given by
%%%%%%
\beq
{\cal G}_{IJ} (\phi^K ) = \frac{ M_{\rm pl}^2}{2 f (\phi^K ) } \left[ \delta_{IJ} + \frac{3}{f (\phi^K ) } f_{, I} f_{, J} \right] ,
\label{GIJ}
\eeq
where $f_{, I} = \partial f / \partial \phi^I$.

We encounter an interesting feature when performing this conformal transformation for models with multiple scalar fields: unlike the well-studied case of a single-field model, in general there does not exist a rescaling of the scalar fields $\phi^I$ that can bring the gravitational portion of the action into Einstein-Hilbert form while also yielding canonical kinetic terms for the scalar fields. In particular, for $M_0 \neq 0$ and $N \geq 2$ scalar fields, the conformal transformation induces a field-space manifold whose metric, ${\cal G}_{IJ} (\phi^K)$, is not conformal to flat \cite{DKconf}.\footnote{In the case of Brans-Dicke-like couplings, with $M_0 = 0$, one may rescale the fields $\phi^I$ to bring ${\cal G}_{IJ} \rightarrow \delta_{IJ}$, and hence restore canonical kinetic terms, only for $N \leq 2$. For $N > 2$, even with $M_0 = 0$, one again finds that ${\cal G}_{IJ}$ is not conformal to flat \cite{DKconf}. } Instead, following the conformal transformation, models within this family assume the form of nonlinear sigma models \cite{Ketov}.

The potential is also stretched by the conformal factor upon transformation to the Einstein frame. In particular, we find
%%%%%%
\beq
V (\phi^I) = \frac{ M_{\rm pl}^4} {[ 2 f (\phi^I) ]^2 } \tilde{V} (\phi^I ) .
\label{Vconf}
\eeq
This is the generalization of Dicke's original finding that masses of particles depend on the Brans-Dicke field following the conformal transformation \cite{Dickeconf}. In the context of simple inflationary models, this conformal stretching of the potential leads to important changes to the inflationary dynamics, compared to models with minimally coupled fields. The most important change is the emergence of strong single-field attractor behavior, which we discuss in Section \ref{sec:SFA}.

Building on pioneering work on multifield inflation \cite{WandsBartolo,Multifield}, we developed in \cite{KMS,GKS,KS,SSK,DeCross} a doubly covariant formalism with which to address dynamics in models that include multiple scalar fields with nonminimal couplings --- that is, covariant with respect to both ordinary gauge transformations ($x^\mu \rightarrow x^{\prime \mu}$) as well as reparameterizations of the field-space coordinates ($\phi^I \rightarrow \phi^{\prime I}$). We consider perturbations around a Friedmann-Lema\^{i}tre-Robertson-Walker spacetime metric, which we take to be spatially flat for convenience; the radius of curvature is stretched exponentially quickly during the first few efolds of inflation, so that a spatially flat background provides an excellent approximation for later dynamics. We then have
%%%%%%%%
\beq
\begin{split}
ds^2 &= g_{\mu\nu} dx^\mu dx^\nu \\
&= - (1 + 2 A) dt^2 + 2a (\partial_i B) dx^i dt + a^2 \left[ (1 - 2 \psi) \delta_{ij} + 2 \partial_i \partial_j E \right] dx^i dx^j ,
\end{split}
\label{ds}
\eeq
where $a (t)$ is the scale factor, and $A (x^\mu), B (x^\mu), \psi (x^\mu)$, and $E(x^\mu)$ characterize the scalar degrees of freedom of the metric perturbations. Given the symmetries of the spacetime, to background order the fields can only depend on time:
%%%%%
\beq
\phi^I (x^\mu) = \varphi^I (t) + \delta \phi^I (x^\mu) .
\label{phivarphi}
\eeq
The magnitude of the velocity vector for the background fields is given by
%%%%
\beq
\vert \dot{\varphi}^I \vert \equiv \dot{\sigma} = \sqrt{ {\cal G}_{IJ} \dot{\varphi}^I \dot{\varphi}^J } ,
\label{sigma}
\eeq
where overdots denote derivatives with respect to cosmic time, $t$. The background fields obey the equation of motion \cite{KMS,Multifield}
%%%%%%%%%%
\beq
{\cal D}_t \dot{\varphi}^I + 3 H \dot{\varphi}^I + {\cal G}^{IJ} V_{, J} = 0 ,
\label{varphieom}
\eeq
where $H \equiv \dot{a} / a$ is the Hubble parameter, and we have introduced a (covariant) directional derivative for vectors $A^I$ on the field-space manifold:
%%%%%%%%%%%
\beq
{\cal D}_t A^I \equiv \dot{\varphi}^J  {\cal D}_J A^I = \dot{A}^I + \Gamma^I_{\> JK} A^J \dot{\varphi}^K .
\label{covD}
\eeq
The Christoffel symbols $\Gamma^I_{\> JK}$ are constructed from the field-space metric ${\cal G}_{IJ}$. The Friedmann equations (to background order) take the form \cite{KMS,Multifield}
%%%%%%%%%
\beq
\begin{split}
H^2 &= \frac{1}{3 M_{\rm pl}^2} \left[ \frac{1}{2} \dot{\sigma}^2 + V (\varphi^I ) \right] , \\
\dot{H} &= - \frac{1}{ 2 M_{\rm pl}^2} \dot{\sigma}^2 .
\end{split}
\label{Friedmann}
\eeq
Eqs. (\ref{varphieom}) and (\ref{Friedmann}) yield self-consistent inflationary solutions, with $\vert \dot{H} \vert \ll H^2$, across wide ranges of coupling constants and initial conditions \cite{KMS,GKS,KS,SSK}. 

The scale of $H$ during inflation is constrained by recent observations. In particular, the present upper bound on the ratio of primordial tensor-to-scalar power spectra, $r$, requires $H_* \leq 3.4 \times 10^{-5} \> M_{\rm pl}$ \cite{Planck}, where the asterisk indicates the value of $H$ at the time when cosmologically relevant perturbations first crossed outside the Hubble radius during inflation. In simple, single-field models of chaotic inflation, one must fine-tune parameters, such as the quartic self-coupling $\lambda \sim {\cal O} (10^{-12})$, in order to accommodate this bound on $H_*$. In models with nonminimal couplings, however, the magnitude of $H_*$ depends on both the Jordan-frame couplings (such as masses, $m_I$, and quartic self-couplings, $\lambda_I$, in $\tilde{V} (\phi^I)$), as well as the nonminimal coupling constants $\xi_I$, due to the conformal stretching of the potential in Eq. (\ref{Vconf}). Hence one may accommodate the observational contraint on $H_*$ without exponentially fine-tuning the parameters \cite{HiggsInflation,KMS,GKS,KS,SSK}.

In order to study the behavior of the fluctuations, we may generalize the gauge-invariant Mukhanov-Sasaki variable to the multifield case, defining a vector of perturbations $Q^I (x^\mu)$ as a linear combination of the field fluctuations, $\delta \phi^I$, and the metric perturbation, $\psi$ \cite{KMS}\footnote{Because the field-space manifold is curved, one must work with a representation of the field fluctuations that is covariant with respect to reparameterizations of the field-space coordinates, as discussed in \cite{KMS} and references therein. That form reduces to Eq. (\ref{Qdef}) to linear order in the field fluctuations, which will suffice for our purposes here.}:
%%%%%%%%%
\beq
Q^I \equiv \delta \phi^I + \frac{\dot{\varphi}^I }{H} \psi .
\label{Qdef}
\eeq
To linear order, the fluctuations $Q^I$ satisfy the equation of motion \cite{KMS,Multifield}
%%%%%%%%%%%%%%
\beq
{\cal D}_t^2 Q^I + 3 H {\cal D}_t Q^I + \left[ \frac{ k^2}{a^2} \delta^I_{\> \>J} + {\cal M}^I_{\> \>J} \right] Q^J = 0 ,
\label{Qeom}
\eeq
where we have performed a Fourier transform, $\nabla^2 Q^I = - k^2 Q^I$ with comoving wavenumber $k$, and the mass-squared matrix is given by
%%%%%%%%%
\beq
{\cal M}^I_{\>\> J} \equiv {\cal G}^{IK} \left( {\cal D}_J {\cal D}_K V \right) - {\cal R}^I_{\>\> LM J} \dot{\varphi}^L \dot{\varphi}^M - \frac{1}{ a^3 M_{\rm pl}^2 } {\cal D}_t \left( \frac{ a^3}{H} \dot{\varphi}^I \dot{\varphi}_J \right) .
\label{MIJ}
\eeq
Here ${\cal R}^I_{\> \>LMJ}$ is the Riemann tensor of the field-space manifold, constructed from ${\cal G}_{IJ}$ (and calculated to background order in the fields, $\varphi^I$); we raise and lower field-space indices with ${\cal G}_{IJ}$. The fluctuations thus acquire three distinct contributions to their effective mass: a term arising from the second derivative of the potential, akin to simple single-field models; a term (proportional to ${\cal R}^I_{\> \>LMJ}$) arising from the curvature of the field-space manifold; and a term (proportional to $1/ M_{\rm pl}^2$) arising from the coupled metric perturbations.

\section{Predictions for Observables}
\label{sec:Observations}

Even to linear order, Eq. (\ref{Qeom}) couples fluctuations $Q^I$ with $Q^J$ and so on. The presence of several interacting degrees of freedom can lead to new observational features in multifield models, with no correlates in simple, single-field models. Two of the most important and best studied examples include the amplification of non-Gaussianities in the primordial power spectrum of curvature perturbations, and the amplification of isocurvature perturbations in addition to adiabatic modes. Non-Gaussianities are generically suppressed in single-field models \cite{Maldacena,XingangReview}, and isocurvature modes do not arise at all in models with only a single scalar degree of freedom \cite{BTW,WandsBartolo}. Given tight constraints on primordial non-Gaussianities and isocurvature perturbations from the latest measurements of the CMB \cite{Planck}, many types of multifield models may therefore be in tension with the latest observations. 

In order to quantify these multifield features, we build on techniques developed in \cite{BTW,WandsBartolo,Multifield} and introduce covariant measures with which to study the perturbation spectra \cite{KMS,GKS,KS,SSK,DeCross}. We introduce a unit vector
%%%%%
\beq
\hat{\sigma}^I \equiv \frac{ \dot{\varphi}^I }{\dot{\sigma} }
\label{hatsigma}
\eeq
which points in the direction of the background fields' evolution. The directions in field space orthogonal to $\hat{\sigma}^I$ are spanned by 
%%%%%%%
\beq
\hat{s}^{IJ} \equiv {\cal G}^{IJ} - \hat{\sigma}^I \hat{\sigma}^J .
\label{hats}
\eeq
We may then project the perturbations $Q^I$ into components along the direction of the background fields' motion (the adiabatic direction) and orthogonal to that motion (the isocurvature directions):
%%%%%%
\beq
Q_\sigma \equiv \hat{\sigma}_I Q^I , \>\>\> \delta s^I \equiv \hat{s}^I_{\> J} Q^J .
\label{QsigmaQs}
\eeq
The gauge-invariant curvature perturbation, ${\cal R}_c$, is defined as \cite{BTW,GaugeInvariant},
%%%%%%%%%%
\beq
{\cal R}_c \equiv \psi - \frac{H}{ ( \rho + p )} \delta q ,
\label{Rcdef}
\eeq
where $\rho$ and $p$ are the background-order energy density and pressure, respectively, and $\delta q$ is the momentum flux of the perturbed fluid, $T^0_{\>\> i} = \partial_i \delta q$. Given the form of the action in Eq. (\ref{SE}), one may show that \cite{KMS}
%%%%%%%%%%
\beq
{\cal R}_c = \frac{H}{\dot{\sigma} } Q_\sigma .
\label{RcQsigma}
\eeq
Primordial curvature perturbations, ${\cal R}_c (x)$, lead to temperature anisotropies in the CMB. Photons that hail from regions of space that had a slightly greater-than-average gravitational potential will be slightly redshifted, upon expending a bit of extra energy to climb out of the potential well, compared to photons from regions of space that had a slightly less-than-average gravitational potential \cite{BTW,GuthKaiser05,GaugeInvariant}. Hence the statistical properties of the tiny temperature anisotropies of the CMB provide a snapshot of primordial inhomogeneities, which in turn help to constrain models of early-universe inflation.

A critical insight \cite{WandsBartolo,Multifield} is that $Q_\sigma$ and $\delta s^I$ are coupled only if the background fields {\it turn} in field space. Hence features like non-Gaussianities and isocurvature perturbations can be amplified in multifield models if the turn-rate, $\omega^I$, is nonvanishing during the late stages of inflation (typically within the last 60 efolds of inflation). The covariant turn-rate may be defined as \cite{KMS}
%%%%%%
\beq
\omega^I \equiv {\cal D}_t \hat{\sigma}^I .
\label{omega}
\eeq
In multifield models, $\omega^I$ need not remain small during inflation, which can amplify features that are not observed in the CMB. 

Consider the limit $\omega^I \rightarrow 0$ first, in which case the perturbations $Q_\sigma$ and $\delta s^I$ remain decoupled. The effective masses for the perturbations take the form \cite{KMS}
%%%%%%%%%
\beq
{\cal M}_{\sigma\sigma} \equiv \hat{\sigma}_I \hat{\sigma}^J {\cal M}^I_{\>\> J} , \>\> \> {\cal M}_{ss} \equiv \hat{s}_I^{\>\>\>\> J} {\cal M}^I_{\>\> J} .
\label{MsigmaMs}
\eeq
In the limit $\vert {\cal M}_{\sigma\sigma} \vert , \vert {\cal M}_{ss} \vert \ll H^2$, each perturbation will evolve during inflation as a (nearly) massless scalar field in (quasi-) de Sitter space, and hence we may expect each perturbation to develop an amplitude of order \cite{BTW,BirrellDavies,Parker}
%%%%%%%%%%
\beq
{\cal P}_{Q} \simeq \left( \frac{H}{2 \pi} \right)^2 ,
\label{PQ}
\eeq
where the power spectrum is defined as ${\cal P}_Q \equiv (2 \pi)^{-2} k^3 \vert Q_\sigma \vert^2$, which we have evaluated for modes of order the Hubble scale, $k \simeq a H$; likewise for ${\cal P}_{ S}$, the power spectrum associated with the conventionally normalized isocurvature perturbations $S^I \equiv (H / \dot{\sigma} ) \delta s^I$. Upon using Eqs. (\ref{Friedmann}), (\ref{RcQsigma}), and the usual definition of the slow-roll parameter,
%%%%%%
\beq
\epsilon \equiv - \frac{ \dot{H} }{ H^2 } ,
\label{epsilon}
\eeq
we therefore expect an amplitude of curvature perturbations during inflation
%%%%%%%%%%%
\beq
{\cal P}_{ R} \simeq \frac{1}{2 M_{\rm pl}^2 \epsilon } \left( \frac{ H}{2 \pi} \right)^2
\eeq
and similarly for ${\cal P}_S$. 

The background fields $\varphi^I (t)$ evolve slowly during inflation, and hence neither $H (t)$ nor $\epsilon (t)$ will remain constant. That means that when modes of various comoving wavenumbers $k$ cross outside the Hubble radius, with $k = aH$, they do so with slightly different amplitudes, ${\cal P}_R (k)$. Hence with a little more work, one may calculate the spectral tilt of the curvature perturbations \cite{BTW,WandsBartolo,Multifield,KMS}:
%%%%%%%%%%%
\beq
n_s \equiv 1 + \frac{ \partial \ln {\cal P}_{ R} }{ \partial \ln k} = 1 - 6 \epsilon + 2 \eta_{\sigma\sigma} ,
\label{ns1}
\eeq
where 
%%%%%%%%
\beq
\eta_{\sigma\sigma} \equiv M_{\rm pl}^2 \frac{ {\cal M}_{\sigma\sigma} }{V}
\label{etasigmasigma}
\eeq
is the generalization of the second slow-roll parameter, for motion along the adiabatic direction. In the limit $\omega^I \rightarrow 0$, therefore, the amplitude and spectral tilt of primordial curvature perturbations in the multifield case look quite similar to the predictions from single-field models --- with one important difference. If $\vert {\cal M}_{ss} \vert \ll H^2$, then such multifield models may amplify a sizable fraction of isocurvature modes, with $\beta_{\rm iso} (k) \equiv {\cal P}_S (k) / [ {\cal P}_{R} (k) + {\cal P}_S (k) ] \sim {\cal O} (1)$ at relevant wavenumbers $k$, which would be in significant tension with the latest observations \cite{Planck}.

Even greater deviations from the single-field case emerge if the fields turn in field space during inflation, $\omega^I \neq 0$. In that case, there can be a transfer of power from the isocurvature to the adiabatic modes, and the amplitude and tilt of ${\cal P}_{ R} (k)$ will be affected. In particular, one may relate the power spectrum at time $t_*$ (say, 50 or 60 efolds before the end of inflation) to its value at some later time, $t$, by means of a transfer function $T_{RS} (t_*, t)$ \cite{KMS,WandsBartolo,Multifield}:
%%%%%%%%%%%
\beq
\begin{split}
{\cal P}_{ R} (k) &= {\cal P}_{ R} (k_*) \left[ 1 + T_{RS}^2 (t_*, t) \right] , \\
n_s  &= n_s (t_*) + \frac{ 1}{H} \left( \frac{ \partial T_{ RS} }{ \partial t_*} \right) \sin \left( 2 \Delta \right) ,
\end{split}
\label{PRnsTRS}
\eeq
where $\Delta \equiv {\rm arccos} (T_{ RS} / \sqrt{ 1 + T_{ RS}^2 } )$. Even a modest transfer of power from the isocurvature to the adiabatic modes could push multifield models out of agreement with the latest high-precision measurements of quantities like $n_s$. Moreover, since $T_{ RS}$ is scale-dependent, such processes effectively couple modes of different wavenumber, $k$, and hence can amplify non-Gaussianities, pushing the coefficient of the bispectrum $f_{NL} \gg {\cal O} (1)$ \cite{KMS,Multifield}.\footnote{To calculate $f_{NL}$ properly, one must go beyond linear order in the fluctuations and calculate the genuine bispectrum, $\langle {\cal R}_c ({\bf k}_1) {\cal R}_c ({\bf k}_2) {\cal R}_c ({\bf k}_3) \rangle$ \cite{KMS,Maldacena,XingangReview}; upon performing the full calculation, we find a strong correlation between nonzero $T_{RS}$ and sizable $f_{NL}$ \cite{KMS}. }

In the Einstein frame, there is no anisotropic pressure to leading order in the perturbations ($\Pi^i_{\> j} \propto T^i_{\> j} \sim 0$ for $i \neq j$), and hence the tensor perturbations $h_{ij}$ evolve just as in single-field models. Around the pivot scale $k_*$, the power spectrum thus obeys ${\cal P}_T \simeq 128 ( H^2 / M_{\rm pl}^2)$ \cite{KS,WandsBartolo,Multifield}, which yields a prediction for the tensor-to-scalar ratio, $r$,
%%%%%%%%
\beq
r \equiv \frac{ {\cal P}_T}{ {\cal P}_R} = \frac{16 \epsilon}{[ 1 + T_{RS}^2] } .
\label{rdef}
\eeq
Just as the case for $n_s$ and $f_{NL}$, predictions for $r$ can deviate strongly from the usual single-field predictions in the case of significant transfer of power from isocurvature to adiabatic modes.

The exact form of $T_{ RS}$ for multifield models with nonminimal couplings may be found in \cite{KMS}; the important point is that $T_{  R S} \propto \vert \omega^I \vert$. In general, when significant turning occurs and $T_{RS} \geq {\cal O} (10^{-1})$, one finds both $n_s$ and $f_{NL}$ pulled significantly outside the $2\sigma$ bounds from the latest observations \cite{KMS,SSK}.

\section{Single-Field Attractor}
\label{sec:SFA}

For multifield models with nonminimal couplings, the turn rate generically remains negligible. Therefore the types of observational consequences that may arise in multifield models, such as the overproduction of isocurvature modes or the amplification of significant non-Gaussianities, typically do not arise for this class of models. The reason comes from the conformal stretching of the potential, $V (\phi^I$) of Eq. (\ref{Vconf}).

For simplicity, consider a two-field model, with $\phi^I = ( \phi, \chi)$. Then for a generic, renormalizable potential in the Jordan frame,
%%%%%%%
\beq
\tilde{V} (\phi, \chi) = \frac{1}{2} m_{\phi}^2 \phi^2 + \frac{1}{2} m_{\chi}^2 \chi^2 + \frac{1}{2} g \phi^2 \chi^2 + \frac{\lambda_\phi}{4} \phi^4 +  \frac{\lambda_\chi}{4} \chi^4 ,
\label{VJ2field}
\eeq
and a nonminimal coupling function $f (\phi, \chi)$ as in Eq. (\ref{fmulti}), we find from Eq. (\ref{Vconf}) the potential in the Einstein frame\footnote{We have set $M_0 = M_{\rm pl}$ in $f (\phi^I)$, since for $\tilde{V} (\phi^I)$ in Eq. (\ref{VJ2field}), the global minimum of the potential occurs at $\phi = \chi = 0$ rather than at any nonzero vacuum expectation value. Hence at the end of inflation, once $\phi$ and $\chi$ settle into the global minimum of the potential, $f (\phi^I) \rightarrow M_{\rm pl}^2 / 2$, recovering the usual gravitational coupling for general relativity.}
%%%%%%
\beq
V (\phi, \chi) = \frac{ M_{\rm pl}^4}{4} \frac{ (2 m_{\phi}^2 \phi^2 + 2 m_\chi^2 \chi^2 + 2 g \phi^2 \chi^2 + \lambda_\phi \phi^4 + \lambda_\chi \chi^4 )}{ [ M_{\rm pl}^2 + \xi_\phi \phi^2 + \xi_\chi \chi^2 ]^2 } .
\label{VE2field}
\eeq
\begin{figure}[t]
\sidecaption
\includegraphics[width=7.5cm]{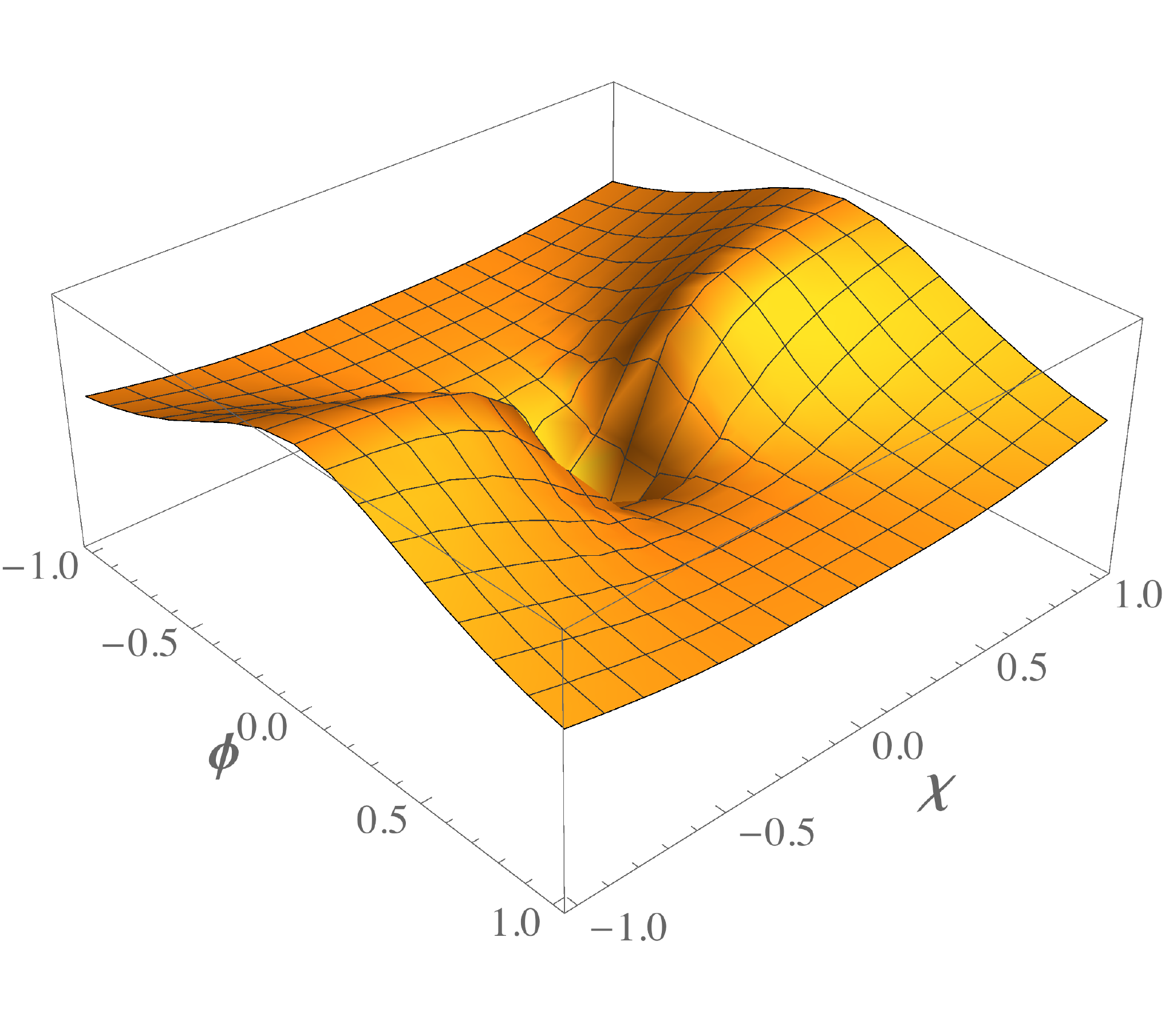}
\caption{The potential $V (\phi, \chi)$ in the Einstein frame, Eq. (\ref{VE2field}), for $\xi_\phi = 100$, $\xi_\chi = 80$, $\lambda_\phi = 10^{-2}$, $\lambda_\chi = 1.25 \times 10^{-2}$, $g = 0.8 \times 10^{-2}$, $m_\phi = 10^{-4} \> M_{\rm pl}$, and $m_\chi = 1.5 \times 10^{-4} \> M_{\rm pl}$. The field values are shown in units of $M_{\rm pl}$.}
\label{fig:1}       % Give a unique label
\end{figure}

\noindent Whereas the potential in the Jordan frame, $\tilde{V} (\phi^I)$, grows as $\phi$ and/or $\chi$ becomes large, in the Einstein frame the potential $V (\phi^I)$ flattens out to long plateaus for large field values. (See Fig. \ref{fig:1}.) That is, generically, the potential in the Einstein frame develops ridges (local maxima) and valleys (local minima), becoming flat along a given direction for asymptotically large field values. Both the ridges and valleys satisfy $V > 0$, and hence the system will inflate (albeit at different rates) whether the fields evolve along a ridge or a valley during inflation. 

\begin{figure}[t]
\sidecaption
\includegraphics[width=7.5cm]{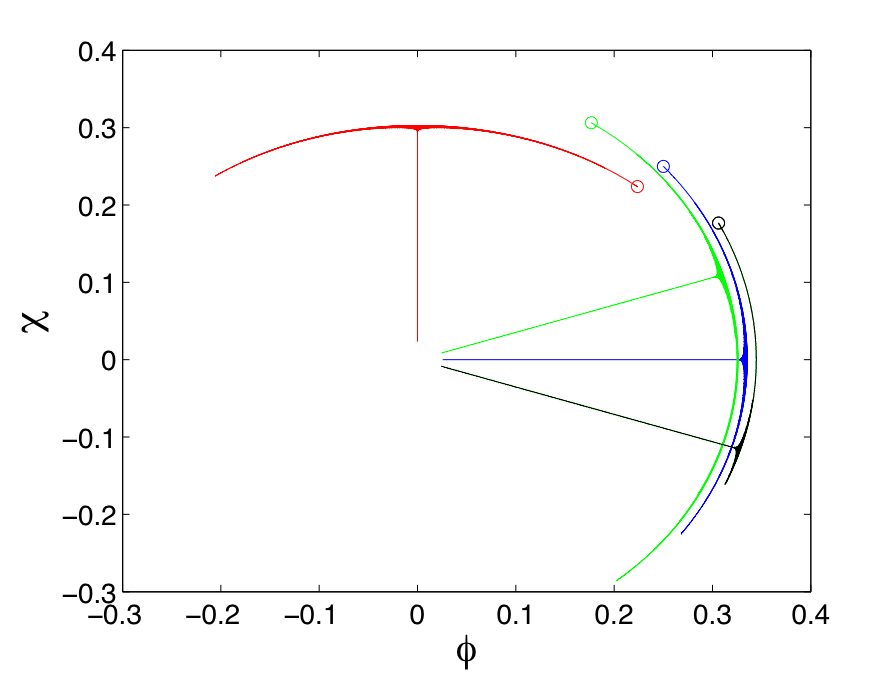}
\caption{Field trajectories for various couplings and initial conditions. Open circles indicate fields' initial values (in units of $M_{\rm pl}$). We set $\dot{\phi}_0 = \dot{\chi}_0 = m_I = 0$, $\xi_\phi = 10^2, \lambda_\phi = 10^{-2}$, and vary the other parameters $\{ \xi_\chi , \lambda_\chi , g , \theta_0 \}$: $\{ 1.2 \xi_\phi, 0.75 \lambda_\phi, \lambda_\phi, \pi/4 \}$ (red); $\{ 0.8 \xi_\phi , \lambda_\phi, \lambda_\phi , \pi / 4 \}$ (blue); $\{0.8 \xi_\phi, \lambda_\phi, 0.75 \lambda_\phi , \pi / 3 \}$ (green); $\{ 0.8 \xi_\phi , 1.2 \lambda_\phi , 0.75 \lambda_\phi , \pi / 6 \}$ (black). Here $\theta_0 \equiv {\rm arctan} (\chi_0 / \phi_0)$. See also \cite{KS,DeCross}.}
\label{fig:2}       % Give a unique label
\end{figure}

The ridge-valley structure of the potential leads to strong single-field attractor behavior during inflation, across a wide range of couplings and initial conditions \cite{KMS,GKS,KS,SSK,DeCross}. If the fields happen to begin evolving along the top of a ridge, they will eventually fall into a neighboring valley at a rate that depends on the local curvature of the potential. Once the fields fall into a valley, Hubble drag quickly damps out any transverse motion in field space, after which the system evolves with virtually no turning for the remainder of inflation. (See Fig. \ref{fig:2}.) In \cite{DeCross}, we demonstrate that the strong attractor behavior persists in the limit $0 < \xi_I \leq 1$ as well as in the limit $\xi_I \gg 1$.

In the limit of strong nonminimal couplings, $\xi_I \gg 1$, the fields rapidly fall into a single-field attractor (within the first few efolds of inflation) unless one fine-tunes the ratio of couplings {\it and} the fields' initial conditions to exponential accuracy. Such attractor behavior is therefore a generic feature of multifield models with nonminimal couplings, and subsumes the class of ``$\alpha$ attractors" that has recently been identified \cite{alpha}.

The lack of turning in field space means that, generically, models in this family yield predictions very similar to those of simple single-field models of ``plateau" inflation. With $\omega^I \simeq 0$, there is virtually no transfer of power from the isocurvature to the adiabatic modes, $T_{RS} \simeq 0$. Moreover, the effective mass of the isocurvature modes, ${\cal M}_{ss}$, remains {\it large} during inflation, while the fields evolve within a valley of the potential: ${\cal M}_{ss} \gg H^2$. Hence $\beta_{\rm iso} (k) \sim {\cal O} (10^{-5})$, well in keeping with recent observational constraints \cite{SSK,Planck}.

Even better, within a single-field attractor and in the limit $\xi_I \gg 1$, one may integrate the equations of motion for the background fields within a slow-roll approximation, taking $\vert \ddot{\varphi}^I  \vert \ll \vert H \dot{\varphi}^I \vert$; as demonstrated in \cite{KS}, the resulting analytic expressions provide a remarkably close match for the exact numerical solutions within a given single-field attractor. In particular, we find \cite{KS}
%%%%%%%%%
\beq
\frac{ \xi_\phi \phi_*^2}{ M_{\rm pl}^2} \simeq \frac{4}{3} N_* ,
\label{Nstarattractor}
\eeq
where $N_*$ is the number of efolds before the end of inflation, and we have considered (in this case) couplings such that the direction $\chi = 0$ is a local minimum of the potential. (We arrive at comparable expressions for other choices of couplings such that the local minimum lies along some angle $\theta = {\rm arctan} (\chi / \phi)$ in field space.) In that limit, we find expressions for the slow-roll parameters that are {\it independent} of the couplings: 
%%%%%%%%%%
\beq
\epsilon \simeq \frac{3}{4 N_*^2} , \>\>\> \eta_{\sigma\sigma} \simeq - \frac{1}{ N_*} \left( 1 - \frac{3}{4 N_*} \right) .
\label{epsilonetastar}
\eeq
Returning to Eqs. (\ref{ns1}), (\ref{PRnsTRS}), and (\ref{rdef}) with $\omega^I \simeq 0$ and hence $T_{RS} \simeq 0$, we then find \cite{KS}
%%%%%%%%%%%
\beq
n_s \simeq 1 - \frac{2}{N_*} - \frac{3}{N_*^2} , \>\>\> r \simeq \frac{12}{N_*^2} ,
\label{ns}
\eeq
independent of the values of the couplings and the fields' initial conditions. For typical reheating scenarios, one expects $50 \leq N_* \leq 60$ to correspond to the time during inflation when perturbations of a given comoving wavenumber first crossed outside the Hubble radius, which later re-entered the Hubble radius around the time the CMB was emitted \cite{ReheatRev}. Selecting $50 \leq N_* \leq 60$ in Eq. (\ref{ns}) yields 
%%%%%%%
\beq
\begin{split}
0.959 &\leq n_s \leq 0.966, \\
0.003 &\leq r \leq 0.005 .
\end{split}
\label{nsrSFA}
\eeq
This value of the spectral index, $n_s$, is in excellent agreement with the latest measurement by the Planck collaboration, $n_s = 0.968 \pm 0.006$ \cite{Planck}, while the predictions for $r$ remain comfortably below the present upper bound of $r < 0.09$ \cite{Planck}. Moreover, predictions for the running of the spectral index, $\alpha = d n_s / d \ln k$, satisfy $\alpha < 10^{-3}$ \cite{KS}, likewise consistent with the latest observational estimates (which themselves are consistent with no observable running) \cite{Planck}. And with $\omega^I \simeq 0$ and hence $T_{RS} \simeq 0$, these models predict $f_{NL} \sim {\cal O} (10^{-1})$ \cite{KMS}, again perfectly consistent with the latest observational constraints \cite{Planck}.

Lastly, one may study post-inflation reheating in this family of models \cite{DeCross}. The single-field attractor persists after the end of inflation, at least during times when the perturbations may be treated to linear order. The lack of turning in field space leads to efficient transfer of energy from the inflaton condensate to coupled fluctuations, in contrast to multifield models with minimal couplings, in which ``dephasing" of the background fields' oscillations typically suppresses resonances \cite{ReheatRev,Barnaby}. Hence reheating in these models should be efficient, with an effective equation of state $w = p / \rho$ that interpolates between $w \simeq 0$ and $w \simeq 1/3$ within the first few efolds after the end of inflation \cite{DeCross}.

\section{Conclusions}
\label{sec:4}

More than half a century after Brans and Dicke introduced their scalar-tensor theory of gravitation, the study of scalar fields with nonminimal couplings continues to flourish. The number of compelling theoretical motivations for considering such nonminimal couplings has grown, and the relevance of such models for understanding the earliest moments of cosmic history is stronger than ever. 

Brans and Dicke introduced their work at a time when Solar System tests of gravitation were still rare, and before the CMB had even been detected! It is an amazing testament to Carl's curiosity and physical insights that work stemming from his dissertation continues to inspire investigations of our cosmos to this day.\footnote{Beyond his work on scalar-tensor gravitation, I also learned of Carl's interest in quantum entanglement and Bell's theorem \cite{BransBell} during my early visits with him  --- work that has also inspired some of my own recent research \cite{cosmicBell}. } Congratulations to Carl on his 80th birthday, with admiration and gratitude.

\begin{acknowledgement}
It is a great pleasure to thank Carl Brans for his kind encouragement over the years. I would also like to thank Torsten Asselmeyer-Maluga for editing this Festschrift in honor of Carl's 80th birthday. I am indebted to Evangelos Sfakianakis for pursuing the research reviewed here with me, along with our collaborators Matthew DeCross, Ross Greenwood, Edward Mazenc, Audrey (Todhunter) Mithani, Anirudh Prabhu, Chanda Prescod-Weinstein, and Katelin Schutz. This research has also benefited from discussions with Bruce Bassett and Alan Guth over the years. This work was conducted in MIT's Center for Theoretical Physics and supported in part by the U.S. Department of Energy under grant Contract Number DE-SC0012567.
\end{acknowledgement}

\end{document}